\documentclass[letterpaper, 12pt]{article}
\usepackage{jheppub}

\usepackage{graphicx}
\usepackage{bbm}
\usepackage{float}
\usepackage{amssymb}
\usepackage{amsthm}
\usepackage{multirow}
\usepackage{physics}
\usepackage{afterpage}
\usepackage{rotating}
\usepackage{upgreek}
\usepackage{amsmath}
\usepackage{mathtools}
\usepackage{epstopdf}
\usepackage{mathrsfs}
\usepackage{scalefnt}
\usepackage[labelformat=simple]{subcaption}
\usepackage{calligra}
\usepackage[T1]{fontenc}  

\begin{document}

\begin{titlepage}

\setcounter{page}{1} \baselineskip=15.5pt \thispagestyle{empty}
{\flushright {YNU-HEP-26-023}\\}
		
\bigskip\
		
\vspace{1.6cm}
\begin{center}
{\Large \bfseries ‌Stochastic Axion Mixing: A General Mechanism\vspace{0.24cm}\\ Beyond Decay Constant Constraints}
\end{center}
\vspace{0.15cm}
			
\begin{center}
{\fontsize{14}{30}\selectfont Hai-Jun Li$^{a,b}$}
\end{center}
\begin{center}
\vspace{0.25 cm}
\textsl{$^a$Department of Physics, Yunnan University, Kunming 650091, China}\\
\textsl{$^b$International Centre for Theoretical Physics Asia-Pacific, Beijing 100190, China}\\

\vspace{-0.1 cm}				
\begin{center}
{E-mail: \textcolor{blue}{\tt {lihaijun@ynu.edu.cn}}}
\end{center}	
\end{center}
\vspace{0.6cm}
\noindent

We propose a novel and generalized mechanism, dubbed stochastic axion mixing. 
In a multi-axion framework, this mixing occurs naturally provided that the masses of all ultra-light axion-like particles (ALPs) are distinct and lighter than the zero-temperature mass of the QCD axion. 
Crucially, this mechanism is independent of the relative magnitudes of the axion decay constants. 
In contrast to the conventional maximal mixing scenario --- which strictly relies on specific decay constant hierarchies --- stochastic mixing represents a significantly broader formalism. 
Notably, maximal mixing emerges as a specific subset of stochastic mixing under restrictive conditions. 
This new mechanism offers profound implications for axion cosmology.
 		
\vspace{3cm}
			
\bigskip
\noindent\today
\end{titlepage}
			
\setcounter{tocdepth}{2}
			
 

\section{Introduction}

It is widely known that a plenitude of axions can originate from higher-dimensional gauge fields \cite{Witten:1984dg, Green:1984sg, Svrcek:2006yi, Conlon:2006tq}. 
We anticipate that one of these axions can solve the strong CP problem in the Standard Model (SM) \cite{Peccei:1977hh, Peccei:1977ur}, while the remaining ones are ultra-light axion-like particles (ALPs). 
Specifically, in the compactifications of type IIB string theory on orientifolds of Calabi-Yau (CY) three-folds, the resulting string axiverse contains one QCD axion and multiple ALPs \cite{Arvanitaki:2009fg, Cicoli:2012sz, Broeckel:2021dpz}.
Additionally, through the misalignment mechanism, these axions are excellent candidates for dark matter (DM) \cite{Preskill:1982cy, Abbott:1982af, Dine:1982ah}.

Axion mixing in the axiverse has recently drawn significant attention \cite{Hill:1988bu, Daido:2015cba, Li:2023uvt, Cyncynates:2023esj, Ho:2018qur, Li:2023xkn, Li:2024jko, Murai:2024nsp, Li:2025cep, Li:2025imm, Murai:2025wbg, Asadi:2025cvm, deGiorgi:2025ldc, Baryakhtar:2026oun}.
In particular, ref.~\cite{Li:2025cep} studied axion mass mixing in the string axiverse and adopted a bottom-up perspective to explore the conditions that an axion model must satisfy to exhibit maximal axion mixing --- the scenario in which the degree of effective mixing reaches its maximum.
Furthermore, the maximal mixing can be classified into the light QCD axion scenario \cite{Daido:2015cba, Li:2023uvt} and the heavy QCD axion scenario \cite{Cyncynates:2023esj}.

\begin{figure}[t]
\centering
\includegraphics[width=0.70\textwidth]{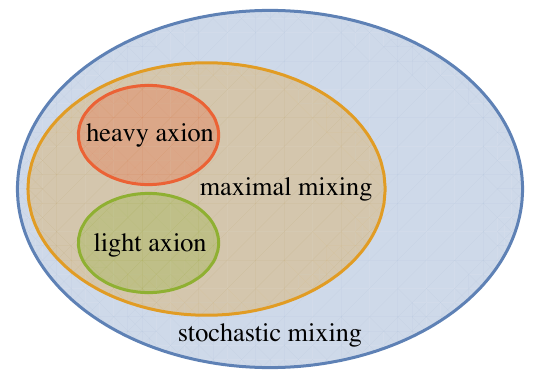}
\caption{Schematic illustration of the scope for different axion mixing scenarios.
Notice that the area sizes here are only for illustrative purposes.}
\label{fig_region}
\end{figure}

In this short letter, we propose a novel axion mixing scenario within the context of multi-axion mass mixing, dubbed stochastic axion mixing. 
Firstly, we briefly review the string axiverse and the maximal axion mixing.
In the canonical mixing scenario, maximal mixing arises under specific conditions: when the masses of all ALPs are smaller than the zero-temperature mass of the QCD axion, with no two ALP masses being identical, and when the decay constants of all ALPs are uniformly either smaller or larger than that of the QCD axion \cite{Li:2025cep}.
However, in the context of stochastic mixing, we find that mixing can occur as long as the masses of all ALPs are unequal and are all smaller than the mass of the QCD axion, and it does not depend on the relation between the axion decay constants. 
We also present a detailed illustration of axion dynamics within the context of stochastic mixing.
Subsequently, we conduct a quantitative analysis of the maximal mixing and stochastic mixing.
We find that stochastic mixing is a more general mixing scenario.
When the decay constants of all ALPs are uniformly either smaller or larger than that of the QCD axion, stochastic mixing encompasses maximal mixing, manifesting as either a light QCD axion scenario or a heavy QCD axion scenario.
See figure~\ref{fig_region} for a schematic illustration of the scope for different axion mixing scenarios.

The rest of this letter is structured as follows.  
In section~\ref{sec_review_axion_mixing}, we briefly review the string axiverse and the maximal axion mixing scenario.
In section~\ref{sec_stochastic_axion_mixing}, we propose a novel and more general stochastic axion mixing scenario.
Subsequently, in section~\ref{sec_quantitative_analysis}, we conduct a quantitative analysis of the maximal mixing and stochastic mixing.
In section~\ref{sec_implications}, we briefly discuss the implications of this new mechanism for axion DM production.
Finally, the conclusion is given in section~\ref{sec_Conclusion}.

\section{Maximal axion mixing}
\label{sec_review_axion_mixing}

In this section, we briefly review the string axiverse and the maximal axion mixing.

\subsection{A plenitude of axions in the string axiverse}

Firstly, we briefly introduce the framework of the string axiverse that encompasses a plenitude of axions.
In the type IIB string axiverse \cite{Cicoli:2012sz}, the QCD axion ($a$) and multiple ultra-light ALPs ($A_i$) can coexist, in which the axion masses and decay constants exhibit a logarithmic distribution rather than a power-law one.

Specifically, the geometric stipulation necessitates the presence of two distinct blow-up modes: one, denoted as $D_{\rm SM}$, is designed to accommodate the SM, while the other, $D_{\rm np}$, is intended to facilitate non-perturbative effects.
The internal volume can be expressed as \cite{Broeckel:2021dpz}
\begin{eqnarray}
\mathcal{V}=\dfrac{1}{6}\sum_{i,j,k=1}^{h^{1,1}-2} k_{ijk} t_i t_j t_k-\gamma_{\rm SM} \tau_{\rm SM}^{3/2}-\gamma_{\rm np}\tau_{\rm np}^{3/2}\, ,
\end{eqnarray} 
where $h^{1,1}$ represents the number of Kähler moduli, $k_{ijk}$ are the intersection numbers, $t_i$ are 2-cycle volumes, $\gamma_i$ are coefficients, and $\tau_i$ are 4-cycle moduli.
The dominant contributions to the scalar potential stem from corrections to Kähler potential $K$ and superpotential $W$. These corrections are responsible for stabilizing three moduli: $\mathcal{V}$, $\tau_{\rm np}$, and $\theta_{\rm np}$. Given that the axion $\theta_{\rm np}$ is excessively massive, it becomes imperative to stabilize the remaining Kähler moduli by integrating additional subleading contributions to both $K$ and $W$.
The resulting axiverse is characterized by a perfect QCD axion $\theta_{\rm SM}$, alongside multiple ultra-light ALPs.
Their masses are given by
\begin{eqnarray}
m_{a,0}\simeq\dfrac{\Lambda_{\rm QCD}^2}{f_a}\, , \quad m_{A_i}\simeq M_p e^{-\pi \tau_i/n_i}\, ,
\end{eqnarray} 
where $\Lambda_{\rm QCD}$ signifies the scale of QCD confinement effects that generate the axion potential, $M_p$ is the Planck mass, and $n_i$ determines the periodicity of the cosine potential.
The corresponding axion decay constants are as follows
\begin{eqnarray}
f_a=\dfrac{c_{\rm SM}}{\tau_{\rm SM}^{1/4}}\dfrac{M_p}{\mathcal{V}^{1/2}}\, , \quad f_{A_i}=\dfrac{c_i}{h_i}\dfrac{M_p}{\mathcal{V}^{2/3}}\, ,
\end{eqnarray} 
where $c_{\rm SM}$ and $c_i$ are coefficients of order unity, and $h_i$ are functions dependent on the intersection numbers and topological quantities.
When contemplating a scenario with the axion count on the order of one hundred, it is advisable to concentrate solely on the region where $\mathcal{V}\gtrsim\mathcal{O}(10^{14})$. 
This approach ensures the attainment of a perfect QCD axion decay constant at the TeV scale.
See also ref.~\cite{Broeckel:2021dpz} for more details.

\subsection{Maximal axion mixing}

In this subsection, we briefly review the maximal axion mixing in the context of multiple axions.
Specifically, it includes one QCD axion ($a$) and $N$ ALPs ($A_i$).
The maximal mixing is defined as the mixing scenario where the axions experience the largest degree of effective mixing.
The low-energy effective Lagrangian that describes the maximal mixing can be formulated as follows \cite{Li:2025cep}
\begin{eqnarray}
\begin{aligned}
\mathcal{L}&\supset\dfrac{1}{2} f_a^2 \left(\partial\theta\right)^2 + \dfrac{1}{2} \sum_{i=1}^N f_{A_i}^2 \left(\partial\Theta_i\right)^2-m_a^2 f_a^2\left[1-\cos\left(n_{00}\theta+\sum_{j=1}^N n_{0j}\Theta_j+\delta_0\right)\right]\\
&-\sum_{i=1}^N m_{A_i}^2 f_{A_i}^2\left[1-\cos\left(n_{i0}\theta+\sum_{j=1}^N n_{ij}\Theta_j+\delta_i\right)\right]\, ,
\end{aligned}
\end{eqnarray}
where $\theta=\phi_0/f_a$ and $\Theta_i=\varphi_i/f_{A_i}$ are the axion angles defined by the axion fields and decay constants, $m_a$ and $m_{A_i}$ represent the axion masses, $n_{ij}$ are the domain wall numbers, and $\delta_i$ are the constant phases which can be assumed to be zero.

Notice that in order to exhibit maximal mixing, two key Conditions must be satisfied: {\bf 1}) the masses of all ALPs cannot be equal and must be smaller than the zero-temperature mass of the QCD axion, and {\bf 2}) the decay constants of all ALPs must simultaneously be either smaller or larger than that of the QCD axion. 
The two different scenarios in Condition {\bf 2} --- being smaller than and larger than --- correspond to the light QCD axion scenario and the heavy QCD axion scenario, respectively. 
In these cases, the matrix of domain wall numbers $n_{ij}$ can be expressed as
\begin{eqnarray}
\mathfrak{n}_{\mathfrak{ij}}=
\left(
\begin{array}{cccccc}
1  & ~ 0 & ~ 0 & ~ 0 &~\cdots & ~ 0\\
1  & ~ 1 & ~ 0 & ~ 0 &~\cdots & ~ 0\\
1  & ~ 0 & ~ 1 & ~ 0 &~\cdots & ~ 0\\
1  & ~ 0 & ~ 0 & ~ 1 &~\cdots & ~ 0\\
\vdots  & ~\vdots & ~\vdots &~\vdots & ~\ddots & ~ \vdots\\
1  & ~ 0 & ~ 0 & ~ 0 & ~ \cdots & ~ 1
\end{array}
\right)\, , \quad
\mathfrak{n}_{\mathfrak{ij}}=
\left(
\begin{array}{cccccc}
1  & ~ 1 & ~ 1 & ~ 1 &~\cdots & ~ 1\\
0  & ~ 1 & ~ 0 & ~ 0 &~\cdots & ~ 0\\
0  & ~ 0 & ~ 1 & ~ 0 &~\cdots & ~ 0\\
0  & ~ 0 & ~ 0 & ~ 1 &~\cdots & ~ 0\\
\vdots  & ~\vdots & ~\vdots &~\vdots & ~\ddots & ~ \vdots\\
0  & ~ 0 & ~ 0 & ~ 0 & ~ \cdots & ~ 1
\end{array}
\right)\, ,
\label{nij_max}
\end{eqnarray}
respectively.
Here $\mathfrak{n}_{\mathfrak{ij}}$ is a $(N+1)\times(N+1)$ matrix, beginning with indices $\mathfrak{i}=0$ and $\mathfrak{j}=0$.
More details about the dynamics and cosmological implications of the maximal axion mixing can be found in ref.~\cite{Li:2025cep}. 

Furthermore, it is important to emphasize that within this context, the $N$ ALPs already satisfy Condition {\bf 2} by default. 
Once Condition {\bf 2} is violated, the maximal mixing will split into two scenarios, each with a reduced number of mixing instances, depending on the relation between the axion decay constants.
Subsequently, we will consider the situation where Condition {\bf 2} is violated.

\section{Stochastic axion mixing}
\label{sec_stochastic_axion_mixing}

In this section, we propose a novel and more general stochastic axion mixing scenario. 
We consider the case where Condition {\bf 2}, as discussed in the previous section, is violated, that is, the decay constants of some ALPs are smaller than that of the QCD axion, while those of the others are larger.
Of course, the case that meets Condition {\bf 2} also applies here. 
Additionally, Condition {\bf 1} still needs to be satisfied in this context.

Specifically, we consider one QCD axion and $P$ ALPs.
Notice that here we use $P$ to represent the number of ALPs in order to draw a comparison with the maximal mixing scenario. 
By default, we assume they satisfy the relation
\begin{eqnarray}
N\leq P\, ,
\end{eqnarray}
and we will discuss this further later.
The low-energy effective potential is given by
\begin{eqnarray}
\begin{aligned}
V&=m_a^2 f_a^2\left[1-\cos\left(n_{00}\theta+\sum_{j=1}^P n_{0j}\Theta_j\right)\right]\\
&+\sum_{i=1}^P m_{A_i}^2 f_{A_i}^2\left[1-\cos\left(n_{i0}\theta+\sum_{j=1}^P n_{ij}\Theta_j\right)\right]\, .
\end{aligned}
\end{eqnarray}
In this case, the matrix of domain wall numbers $n_{ij}$ is taken as follows
\begin{eqnarray}
\mathfrak{n}_{\mathfrak{ij}}=
\left(
\begin{array}{cccccc}
1  & ~ d_1 & ~ d_2 & ~ d_3 &~\cdots & ~ d_P\\
\tilde{d_1} & ~ 1 & ~ 0 & ~ 0 &~\cdots & ~ 0\\
\tilde{d_2}  & ~ 0 & ~ 1 & ~ 0 &~\cdots & ~ 0\\
\tilde{d_3}  & ~ 0 & ~ 0 & ~ 1 &~\cdots & ~ 0\\
\vdots  & ~\vdots & ~\vdots &~\vdots & ~\ddots & ~ \vdots\\
\tilde{d}_P  & ~ 0 & ~ 0 & ~ 0 & ~ \cdots & ~ 1
\end{array}
\right)\, ,
\label{nij_new}
\end{eqnarray}
where $\tilde{d_i}=1-d_i$, and $d_i$ are defined by
\begin{eqnarray}
d_i=
\begin{cases}
0\,, &\text{if } f_a\gg f_{A_i}\\
1\,, &\text{if } f_a\ll f_{A_i}
\end{cases}
\end{eqnarray}
corresponding to the previous light and heavy QCD axion scenario conditions, respectively.
The equations of motion (EOMs) are given by
\begin{eqnarray}
\ddot\phi_0+3H\dot\phi_0+ m_a^2 f_a\sin\left(\dfrac{\phi_0}{f_a}+\sum_{i=1}^P\dfrac{d_i \varphi_i}{f_{A_i}}\right)+ \sum_{i=1}^P\dfrac{\tilde{d_i}m_{A_i}^2 f_{A_i}^2}{f_a}\sin\left(\dfrac{\tilde{d_i}\phi_0}{f_a}+\dfrac{\varphi_i}{f_{A_i}}\right)=0 \, ,~ 
\end{eqnarray}
\begin{eqnarray}
\ddot\varphi_1+3H\dot\varphi_1+ \dfrac{d_1 m_a^2 f_a^2}{f_{A_1}}\sin\left(\dfrac{\phi_0}{f_a}+\sum_{i=1}^P\dfrac{d_i \varphi_i}{f_{A_i}}\right)+ m_{A_1}^2 f_{A_1}\sin\left(\dfrac{\tilde{d_i}\phi_0}{f_a}+\dfrac{\varphi_1}{f_{A_1}}\right)=0\, , 
\end{eqnarray}
\begin{eqnarray}
\ddot\varphi_2+3H\dot\varphi_2+ \dfrac{d_2 m_a^2 f_a^2}{f_{A_2}}\sin\left(\dfrac{\phi_0}{f_a}+\sum_{i=1}^P\dfrac{d_i \varphi_i}{f_{A_i}}\right)+ m_{A_2}^2 f_{A_2}\sin\left(\dfrac{\tilde{d_i}\phi_0}{f_a}+\dfrac{\varphi_2}{f_{A_2}}\right)=0\, , 
\end{eqnarray}
$\hspace{7cm}\vdots$
\begin{eqnarray}
\ddot\varphi_P+3H\dot\varphi_P+ \dfrac{d_P m_a^2 f_a^2}{f_{A_P}}\sin\left(\dfrac{\phi_0}{f_a}+\sum_{i=1}^P\dfrac{d_i \varphi_i}{f_{A_i}}\right)+ m_{A_P}^2 f_{A_P}\sin\left(\dfrac{\tilde{d}_P\phi_0}{f_a}+\dfrac{\varphi_P}{f_{A_P}}\right)=0\, .~~ 
\end{eqnarray}
Provided that the oscillation amplitudes of axion fields are substantially smaller than their corresponding decay constants, we can approximate the system and derive the mass mixing matrix within this model
\begin{eqnarray}
\mathbf{M}^2=
\left(
\begin{array}{cccc}
m_a^2+\dfrac{1}{f_a^2} \displaystyle\sum_{i=1}^P\tilde{d_i}^2 m_{A_i}^2 f_{A_i}^2& \dfrac{d_1 m_a^2 f_a^2+\tilde{d_1}m_{A_1}^2 f_{A_1}^2}{f_a f_{A_1}} & \cdots & \dfrac{d_P m_a^2 f_a^2+\tilde{d}_Pm_{A_P}^2 f_{A_P}^2}{f_a f_{A_P}}\\
\dfrac{d_1 m_a^2 f_a^2+\tilde{d_1}m_{A_1}^2 f_{A_1}^2}{f_a f_{A_1}}   & m_{A_1}^2+\dfrac{d_1^2 m_a^2 f_a^2}{f_{A_1}^2} & \cdots & \dfrac{d_1 d_P m_a^2 f_a^2}{f_{A_1} f_{A_P}}\\
\vdots  & \vdots & \ddots & \vdots\\
\dfrac{d_P m_a^2 f_a^2+\tilde{d}_Pm_{A_P}^2 f_{A_P}^2}{f_a f_{A_P}} & \dfrac{d_1 d_P m_a^2 f_a^2}{f_{A_1} f_{A_P}} & \cdots & m_{A_P}^2+\dfrac{d_P^2 m_a^2 f_a^2}{f_{A_P}^2} 
\end{array}
\right) .~~~~
\end{eqnarray}
Given our Condition {\bf 1} that the masses of ALPs are all unequal, the mass mixing matrix in this context can be represented as multiple effective matrices
\begin{eqnarray}
\mathbf{M}_i^2=
\left(\begin{array}{cc}
m_a^2+\dfrac{\tilde{d_i}^2 m_{A_i}^2 f_{A_i}^2}{f_a^2} & ~ \dfrac{d_i m_a^2 f_a^2+\tilde{d_i}m_{A_i}^2 f_{A_i}^2}{f_a f_{A_i}}\\
\dfrac{d_i m_a^2 f_a^2+\tilde{d_i}m_{A_i}^2 f_{A_i}^2}{f_a f_{A_i}} & ~ m_{A_i}^2+\dfrac{d_1^2 m_a^2 f_a^2}{f_{A_i}^2}
\end{array}\right)\, ,
\end{eqnarray} 
with the corresponding mass eigenvalues
\begin{eqnarray}
\begin{aligned}
m_{h_i,l_i}^2&=\dfrac{1}{2}\left[m_a^2+m_{A_i}^2+\dfrac{d_i^2 m_a^2 f_a^2}{f_{A_i}^2}+\dfrac{\tilde{d_i}^2 m_{A_i}^2 f_{A_i}^2}{f_a^2}\right]\\
&\pm\dfrac{1}{2 f_a^2 f_{A_i}^2}\bigg[-4\left(1-d_i \tilde{d_i} \right)^2 m_a^2 m_{A_i}^2 f_a^4 f_{A_i}^4\\
&+\bigg(\left(m_a^2+ m_{A_i}^2\right)f_a^2 f_{A_i}^2+d_i^2 m_a^2 f_a^4+\tilde{d_i}^2 m_{A_i}^2 f_{A_i}^4\bigg)^2\bigg]^{1/2}\, .
\end{aligned}
\end{eqnarray} 
Then the effective mass eigenvalues $m_{e_i}$ are given by
\begin{eqnarray}
m_{e_1}&=&m_{h_P}\, ,\\
m_{e_2}&\simeq&
\begin{cases}
m_{l_P}\,, &\text{if } T \le T_{\times_{P-1}}^{(m)}\\
m_{h_{P-1}}\,, &\text{if } T > T_{\times_{P-1}}^{(m)}
\end{cases}\\
m_{e_3}&\simeq&
\begin{cases}
m_{l_{P-1}}\, ,& \text{if } T \le T_{\times_{P-2}}^{(m)}\\
m_{h_{P-2}}\, ,& \text{if } T > T_{\times_{P-2}}^{(m)}
\end{cases}\\
&\vdots\nonumber\\
m_{e_P}&\simeq&
\begin{cases}
m_{l_2}\, , &\text{if } T \le T_{\times_1}^{(m)}\\
m_{h_1}\, , &\text{if } T > T_{\times_1}^{(m)}
\end{cases}\\
m_{e_{P+1}}&=&m_{l_1}\, ,
\end{eqnarray}
where $T_{\times_i}^{(m)}$ represent the intermediate temperatures between the mass crossing temperatures, and the precise values of these temperatures need not be specified.

Notice that the matrix of domain wall numbers in eq.~\eqref{nij_new} combines the two cases considered in eq.~\eqref{nij_max} into one situation. 
Due to the inherently stochastic nature of domain wall numbers within the matrix $\mathfrak{n}_{\mathfrak{ij}}$, we thus refer to this new mixing scenario as {\it stochastic axion mixing}. 
To elaborate, when considering the scenario with $P$ ALPs, stochastic mixing can occur as long as Condition {\bf 1} is satisfied.
This is different from the situation in the maximal mixing scenario, so we have the relation $N\leq P$, and the equality holds when Condition {\bf 2} is further satisfied.
Consequently, our stochastic mixing represents a more general framework for axion mixing. 

\section{Quantitative analysis of maximal mixing \& stochastic mixing}
\label{sec_quantitative_analysis}

In this section, we conduct a quantitative analysis of the maximal mixing and stochastic mixing.
An analysis of the maximal mixing can be found in ref.~\cite{Li:2025cep}.

Quantitatively, we can associate these mixing scenarios with the number of times the single mass crossing phenomenon occurs $n_\times$.
In the maximal mixing scenario, if $N$ ALPs satisfy Conditions {\bf 1} and {\bf 2}, maximal mixing can occur, and we have $n_\times=N$.
If only Condition {\bf 2} is violated, the maximal mixing will be split into two light and heavy axion scenarios, corresponding to $n_\times=a$ and $n_\times=b$ respectively, with $a+b=N$. 
However, in the stochastic mixing scenario, as long as $P$ ALPs satisfy Condition {\bf 1}, regardless of whether Condition {\bf 2} is met, stochastic mixing can take place, and we have the relation
\begin{eqnarray}
n_\times=P\, .
\end{eqnarray}
Notice that this equation always holds true for stochastic mixing.

Here we demonstrate how a given model exhibits stochastic mixing.
Notice that Condition {\bf 1} is still assumed to hold by default, and thus we focus solely on the relation between the axion decay constants and conduct a general analysis.
To simplify the discussion, we examine the scenario where the QCD axion and ten ALPs are mixed together. 
We use $\eta_i$ to represent the ratio between the decay constant of the ALP and that of the QCD axion. 
Take an example, we look into a situation in which $\log_{10}\left(\eta_i\right)$ follows a random distribution spanning the interval $\left(-3,\, 3\right)$,
\begin{eqnarray}
\{-1.65,1.4,0.67,0.59,-0.31,0.93,0.53,-0.96,1.65,-2.4\}\, .
\end{eqnarray}
We assume that these ALPs $A_i$ are arranged in ascending order of mass from smallest to largest.
In the context of the canonical mixing framework, maximal mixing cannot occur; rather, it is divided into two mixing scenarios: $1+4$ mixing and $1+6$ mixing with $n_\times=4$ and $n_\times=6$, respectively.
Then we have the matrices of domain wall numbers
\begin{eqnarray}
\mathfrak{n}_{5\times5}=
\left(
\begin{array}{ccccc}
1  & ~ 0 & ~ 0 & ~ 0 & ~ 0\\
1  & ~ 1 & ~ 0 & ~ 0 & ~ 0\\
1  & ~ 0 & ~ 1 & ~ 0 & ~ 0\\
1  & ~ 0 & ~ 0 & ~ 1 & ~ 0\\
1  & ~ 0 & ~ 0 & ~ 0 &  ~ 1\\
\end{array}
\right)\, , \quad 
\mathfrak{n}_{7\times7}=
\left(
\begin{array}{ccccccc}
1  & ~ 1 & ~ 1 & ~ 1 &~1 & ~ 1& ~ 1\\
0  & ~ 1 & ~ 0 & ~ 0 &~0 & ~ 0& ~ 0\\
0  & ~ 0 & ~ 1 & ~ 0 &~0 & ~ 0& ~ 0\\
0  & ~ 0 & ~ 0 & ~ 1 &~0 & ~ 0& ~ 0\\
0  & ~ 0 & ~ 0 & ~ 0 &~1 & ~ 0& ~ 0\\
0  & ~ 0 & ~ 0 & ~ 0 &~0 & ~ 1& ~ 0\\
0  & ~ 0 & ~ 0 & ~ 0 &~0 & ~ 0& ~ 1
\end{array}
\right)\, ,
\end{eqnarray}
corresponding to the light and heavy QCD axion scenarios in eq.~\eqref{nij_max}, respectively.
In this case, the explicit forms of the effective mixing potential‌ are given by
\begin{eqnarray}
\begin{aligned}
V_{\rm light}&=m_a^2 f_a^2\left[1-\cos\left(\dfrac{\phi_0}{f_a}\right)\right]+m_{A_1}^2 f_{A_1}^2\left[1-\cos\left(\dfrac{\phi_0}{f_a}+\dfrac{\varphi_1}{f_{A_1}}\right)\right]\\
&+m_{A_5}^2 f_{A_5}^2\left[1-\cos\left(\dfrac{\phi_0}{f_a}+\dfrac{\varphi_5}{f_{A_5}}\right)\right]+m_{A_8}^2 f_{A_8}^2\left[1-\cos\left(\dfrac{\phi_0}{f_a}+\dfrac{\varphi_8}{f_{A_8}}\right)\right]\\
&+m_{A_{10}}^2 f_{A_{10}}^2\left[1-\cos\left(\dfrac{\phi_0}{f_a}+\dfrac{\varphi_{10}}{f_{A_{10}}}\right)\right]\, ,
\end{aligned}
\end{eqnarray}
and 
\begin{eqnarray}
\begin{aligned}
V_{\rm heavy}&=m_a^2 f_a^2\left[1-\cos\left(\dfrac{\phi_0}{f_a}+\dfrac{\varphi_2}{f_{A_2}}+\dfrac{\varphi_3}{f_{A_3}}+\dfrac{\varphi_4}{f_{A_4}}+\dfrac{\varphi_6}{f_{A_6}}+\dfrac{\varphi_7}{f_{A_7}}+\dfrac{\varphi_9}{f_{A_9}}\right)\right]\\
&+m_{A_2}^2 f_{A_2}^2\left[1-\cos\left(\dfrac{\varphi_2}{f_{A_2}}\right)\right]+m_{A_3}^2 f_{A_3}^2\left[1-\cos\left(\dfrac{\varphi_3}{f_{A_3}}\right)\right]\\
&+m_{A_4}^2 f_{A_4}^2\left[1-\cos\left(\dfrac{\varphi_4}{f_{A_4}}\right)\right]+m_{A_6}^2 f_{A_6}^2\left[1-\cos\left(\dfrac{\varphi_6}{f_{A_6}}\right)\right]\\
&+m_{A_7}^2 f_{A_7}^2\left[1-\cos\left(\dfrac{\varphi_7}{f_{A_7}}\right)\right]+m_{A_9}^2 f_{A_9}^2\left[1-\cos\left(\dfrac{\varphi_9}{f_{A_9}}\right)\right]\, .
\end{aligned}
\end{eqnarray} 
Nevertheless, if the stochastic mixing scenario proposed in this work is taken into account, the matrix of domain wall numbers corresponding to eq.~\eqref{nij_new} is expressed as
\begin{eqnarray}
\mathfrak{n}_{11\times11}=
\left(
\begin{array}{ccccccccccc}
1  & ~ 0 & ~ 1 & ~ 1 &~1 & ~ 0& ~ 1& ~ 1& ~ 0& ~ 1& ~ 0\\
1  & ~ 1 & ~ 0 & ~ 0 &~0 & ~ 0& ~ 0& ~ 0& ~ 0& ~ 0& ~ 0\\
0  & ~ 0 & ~ 1 & ~ 0 &~0 & ~ 0& ~ 0& ~ 0& ~ 0& ~ 0& ~ 0\\
0  & ~ 0 & ~ 0 & ~ 1 &~0 & ~ 0& ~ 0& ~ 0& ~ 0& ~ 0& ~ 0\\
0  & ~ 0 & ~ 0 & ~ 0 &~1 & ~ 0& ~ 0& ~ 0& ~ 0& ~ 0& ~ 0\\
1  & ~ 0 & ~ 0 & ~ 0 &~0 & ~ 1& ~ 0& ~ 0& ~ 0& ~ 0& ~ 0\\
0  & ~ 0 & ~ 0 & ~ 0 &~0 & ~ 0& ~ 1& ~ 0& ~ 0& ~ 0& ~ 0\\
0  & ~ 0 & ~ 0 & ~ 0 &~0 & ~ 0& ~ 0& ~ 1& ~ 0& ~ 0& ~ 0\\
1  & ~ 0 & ~ 0 & ~ 0 &~0 & ~ 0& ~ 0& ~ 0& ~ 1& ~ 0& ~ 0\\
0  & ~ 0 & ~ 0 & ~ 0 &~0 & ~ 0& ~ 0& ~ 0& ~ 0& ~ 1& ~ 0\\
1  & ~ 0 & ~ 0 & ~ 0 &~0 & ~ 0& ~ 0& ~ 0& ~ 0& ~ 0& ~ 1
\end{array}
\right)
\, .
\end{eqnarray}
In this case, we have $n_\times=10$.
That is to say, in the canonical mixing scenario, the number of times mass crossing occurs is equal to or less than the number of ALPs, with $n_\times \leq N$.
Maximal mixing can only take place when Condition {\bf 2} is satisfied, that is, $n_\times= N$. 
However, in the stochastic mixing scenario, the number of times mass crossing occurs is always equal to the number of ALPs, with $n_\times=P$, regardless of whether Condition {\bf 2} is met. 
This represents the essential difference between them.
In addition‌, the new effective mixing potential‌ is given by
\begin{eqnarray}
\begin{aligned}
V&=m_a^2 f_a^2\left[1-\cos\left(\dfrac{\phi_0}{f_a}+\dfrac{\varphi_2}{f_{A_2}}+\dfrac{\varphi_3}{f_{A_3}}+\dfrac{\varphi_4}{f_{A_4}}+\dfrac{\varphi_6}{f_{A_6}}+\dfrac{\varphi_7}{f_{A_7}}+\dfrac{\varphi_9}{f_{A_9}}\right)\right]\\
&+m_{A_1}^2 f_{A_1}^2\left[1-\cos\left(\dfrac{\phi_0}{f_a}+\dfrac{\varphi_1}{f_{A_1}}\right)\right]+m_{A_2}^2 f_{A_2}^2\left[1-\cos\left(\dfrac{\varphi_2}{f_{A_2}}\right)\right]\\
&+m_{A_3}^2 f_{A_3}^2\left[1-\cos\left(\dfrac{\varphi_3}{f_{A_3}}\right)\right]+m_{A_4}^2 f_{A_4}^2\left[1-\cos\left(\dfrac{\varphi_4}{f_{A_4}}\right)\right]\\
&+m_{A_5}^2 f_{A_5}^2\left[1-\cos\left(\dfrac{\phi_0}{f_a}+\dfrac{\varphi_5}{f_{A_5}}\right)\right]+m_{A_6}^2 f_{A_6}^2\left[1-\cos\left(\dfrac{\varphi_6}{f_{A_6}}\right)\right]\\
&+m_{A_7}^2 f_{A_7}^2\left[1-\cos\left(\dfrac{\varphi_7}{f_{A_7}}\right)\right]+m_{A_8}^2 f_{A_8}^2\left[1-\cos\left(\dfrac{\phi_0}{f_a}+\dfrac{\varphi_8}{f_{A_8}}\right)\right]\\
&+m_{A_9}^2 f_{A_9}^2\left[1-\cos\left(\dfrac{\varphi_9}{f_{A_9}}\right)\right]+m_{A_{10}}^2 f_{A_{10}}^2\left[1-\cos\left(\dfrac{\phi_0}{f_a}+\dfrac{\varphi_{10}}{f_{A_{10}}}\right)\right]\, .
\end{aligned}
\end{eqnarray}  

\section{Implications for axion dark matter production}
\label{sec_implications}

This stochastic mixing mechanism holds the potential to unveil new insights and implications for axion cosmology, especially when embedded in the recently proposed multi-component axion DM framework \cite{Li:2025uwq}.

In this multi-component scenario, DM comprises the QCD axion and multiple ultra-light ALPs. 
While previous studies typically employed a maximal mixing framework, incorporating the stochastic mixing scheme proposed here reveals novel dynamics in axion production.
On the one hand, since stochastic mixing relaxes constraints on axion decay constants, scenarios previously incompatible with maximal mixing can now occur naturally. 
This significantly expands the landscape of viable theoretical axion models. 
Leveraging this mechanism, we can more naturally address both the overproduction of QCD axion DM and the underproduction of QCD axion DM in the dark dimension scenario.
On the other hand, we observe that the stochastic mixing scheme entails a larger population of ALPs. 
Given that these ALPs possess masses below the zero-temperature mass of the QCD axion and exhibit a hierarchical distribution, a significant fraction are inherently ultra-light.
Notably, those with smaller decay constants hold promise for detection in future axion experiments.

While we have primarily addressed the direct impacts of stochastic mixing on axion DM production, other crucial cosmological consequences remain to be explored. 
These include implications for axion relic density, isocurvature fluctuations, dark energy, domain walls, gravitational waves, and primordial black holes, offering promising directions for future work.

\section{Conclusion}
\label{sec_Conclusion}

In summary, we have proposed stochastic axion mixing, a generalized mechanism that relaxes the strict decay constant constraints imposed by the conventional maximal mixing framework. 
Unlike its predecessor, this new scenario requires only that ALP masses be unequal and lighter than the QCD axion, rendering the mixing process independent of the relative magnitudes of decay constants. 
Consequently, maximal mixing is identified merely as a specific subset of stochastic mixing, applicable only under restrictive hierarchical conditions.

Our quantitative analysis confirms that stochastic mixing guarantees a mass crossing for every ALP ($n_\times=P$). 
This ensures maximal mixing efficiency regardless of specific parameters, a robustness that significantly broadens the landscape of viable string axiverse models compared to the limited mass crossing ($n_\times \leq N$) allowed in maximal mixing.

The cosmological implications of this mechanism are profound. 
By relaxing theoretical constraints, stochastic mixing provides a natural solution to the QCD axion DM overproduction problem and addresses abundance deficits in the dark dimension scenario.
Furthermore, it predicts a richer spectrum of ultra-light ALPs, thereby enhancing the prospects for detection in upcoming experiments.

In our upcoming work, we will extend this framework to explore its impact on gravitational waves and primordial black holes, opening new avenues for probing the string axiverse.

\section*{Acknowledgments}

This work was supported in part by Yunnan University and the International Centre for Theoretical Physics Asia-Pacific.




\bibliographystyle{JHEP}
\bibliography{references}

\end{document}